\documentclass[epj-spec]{svjour}

\usepackage{graphics}
\usepackage{amsmath}
\usepackage{amssymb}

\begin{document}

\title{Charge density wave in graphene:\\
magnetic-field-induced Peierls instability}
\author{Jean-No\"{e}l
Fuchs\inst{1}\fnmsep\thanks{\email{fuchs@lps.u-psud.fr}}
\and Pascal Lederer\inst{1} } \institute{Laboratoire de Physique des
Solides, Univ. Paris-Sud, CNRS, UMR 8502, F-91405 Orsay Cedex,
France}

\abstract{We suggest that a magnetic-field-induced Peierls
instability accounts for the recent experiment of Zhang \emph{et
al.} in which unexpected quantum Hall plateaus were observed at high
magnetic fields in graphene on a substrate. This Peierls instability
leads to an out-of-plane lattice distortion resulting in a charge
density wave (CDW) on sublattices A and B of the graphene honeycomb
lattice. We also discuss alternative microscopic scenarios proposed
in the literature and leading to a similar CDW ground state in
graphene.}

\maketitle

\section{Introduction}
\label{intro} The recent interest in graphene was triggered by the
observation of an unusual ``relativistic'' quantum Hall (QH) effect
by two groups \cite{Novoselov,Zhang1}. This effect is a direct
consequence of the peculiar density of states of graphene close to
half-filling \cite{SchakelZhengGusyninPeres}. Shortly after these
pioneering experiments, Zhang et al. \cite{Zhang2} explored the
regime of even larger magnetic fields, up to 45~T, and observed
extra plateaus in the Hall conductivity $\sigma_{xy}$. In contrast
to the relativistic quantum Hall effect observed at lower magnetic
fields, these extra plateaus cannot be explained by considering
non-interacting electrons. In the present paper, we review the
recent proposal of a magnetic-field-induced Peierls instability for
graphene on a substrate \cite{Fuchs}, which gives an explanation to
the extra quantum Hall plateaus observed at high magnetic field. The
Peierls instability leads to an out-of-plane lattice distortion
resulting in a charge density wave (CDW) on sublattices A and B of
the graphene honeycomb lattice. Other mechanisms leading to such a
CDW ground state in graphene have been proposed in the literature
and are discussed in the present paper
\cite{Khveshchenko,Gusynin,Herbut,Alicea}. The originality of the
magnetic-field-induced Peierls instability scenario \cite{Fuchs} is
to rely on electron-phonon coupling via the substrate on which the
graphene flake lies, rather than on Coulomb interactions between
electrons \cite{Khveshchenko,Gusynin,Herbut,Alicea}.

\section{Quantum Hall effect in graphene: non-interacting electrons}
\label{QHEgraphene} The graphene  crystal is a honeycomb lattice of
carbon atoms: a two-dimensional triangular Bravais lattice with a
basis of two atoms, usually referred to as A and B. The distance
between nearest neighbor atoms is $a=0.14$~nm and the lattice
constant is $a\sqrt{3}$. Experimentally, graphene sheets of area
$\mathcal{A} \sim (3 - 10\, \mu$m$)^2$ are deposited on amorphous
SiO$_2$/Si substrate. Applying a gate voltage $V_g$ via the
substrate allows one to control the electronic filling of the
graphene bands, which amounts to doping. The number of induced
electronic charges is given by $N_c=V_g C_g/e$ where the capacitance
per unit area can be estimated as $C_g/\mathcal{A}\approx \epsilon
/d \approx 1.2\times 10^{-4}$~F/m$^2$, where $-e<0$ is the electron
charge, $\epsilon=\epsilon_0 \epsilon_r$ (with $\epsilon_r \approx
4$) is the silicon oxide dielectric constant and its thickness is
$d\sim 300$~nm \cite{Novoselov,Zhang1}.

The electronic properties of graphene are described by a standard
nearest-neighbor tight-binding model \cite{Wallace} with hopping
amplitude $t\approx 3$~eV \cite{Saito} between $2p_z$ carbon
orbitals. Each carbon atom contributes one conduction electron.
There are $2N_p$ electrons in the sample under zero gate voltage,
where $N_p$ is the number of unit cells in the sample. The first
Brillouin zone is hexagonal. Among its six corners, only two are
inequivalent and called $K$ and $K'$. We choose
$\mathbf{K}=4\pi/(3\sqrt{3}a)\mathbf{u}_x$ and
$\mathbf{K'}=-\mathbf{K}$. The resulting band structure features the
merging of the conduction and valence band at precisely these two
points: graphene is a two-valley ($K$ and $K'$) zero-gap
semiconductor. Near these so-called Dirac points, the electrons
behave as charged massless chiral fermions and the valence and
conduction bands have the shape of a diabolo. The relativistic-like
dispersion relation of electrons is $\varepsilon_k=\pm \hbar v_F
|\mathbf{k}|$, with a Fermi velocity $v_F=3at/2\hbar \approx
10^6$~m/s about 300 times smaller than the velocity of light $c$.
This results in a linear density of states close to the Dirac point.
In undoped graphene ($V_g=0$), the valence band is filled and the
conduction band is empty: the chemical potential is right at the
Dirac points. Changing the gate voltage from zero allows one to fill
the conduction band or empty the valence band.

In order to discuss the quantum Hall effect in graphene, we briefly
review the Landau problem of a graphene electron in a perpendicular
magnetic field \cite{McClure}. The magnetic field is assumed weak
enough to neglect any magnetic flux commensuration effect (for a
study of the Hofstadter butterfly in graphene, see \cite{Rammal})
and we use the continuum approximation in which the tight-binding
model reduces to the two-dimensional massless Dirac Hamiltonian. The
Landau levels (LL) are
\begin{equation}
\varepsilon_{n}=\text{sgn}(n)\sqrt{|n|}\hbar \omega_c\, ,
\label{LL1}
\end{equation}
where the ``cyclotron'' frequency is $\omega_c=v_F
\sqrt{2eB_\perp/\hbar}$, and the LL index $n$ is an integer. The
unusual LL structure with a square root dependence in the
perpendicular magnetic field is a direct consequence of the linear
density of states close to the Dirac point. Each LL is $4N_\phi$
times degenerate, where the number of flux quanta across the sample
$N_\phi=e B_\perp \mathcal{A}/(2\pi \hbar)$ gives the orbital
degeneracy and the factor $4$ accounts for twofold spin $1/2$ and
twofold valley degeneracy.

Let $\nu=N_c/N_\phi=C_g V_g/e N_\phi$ be the filling factor. When
the gate voltage is zero, so is the filling factor and the central
Landau level (CLL, $n=0$) is half-filled as a result of
particle-hole symmetry. In addition, each LL is $4N_\phi$ times
degenerate. These two facts alone show that a plateau at $\nu e^2/h$
occurs in the Hall conductivity $\sigma_{xy}$ each time the filling
factor is close to $\text{sgn}(n)(4|n|+2)=\pm2;\pm6;\pm10;\dots$. In
order to have extra quantum Hall plateaus, one has to provide
mechanisms for lifting the spin and valley degeneracy. In the
absence of interactions between electrons, the only such mechanism
is the Zeeman effect, which lifts the spin degeneracy, but does not
affect the twofold valley degeneracy. The Zeeman effect, which
depends on the total magnetic field $B_\text{tot}$, splits each LL
and leads to a gap $\Delta_Z=g^* \mu_B B_\text{tot}$, where
$\mu_B=e\hbar/2m$ is the Bohr magneton, $m$ is the bare electron
mass, and $g^*\approx 2$ is the effective $g$-factor \cite{Zhang2}.
If the Zeeman gap is larger than the LL width, one expects extra
quantum Hall plateaus at $\nu=4n=0;\pm4;\pm8;\dots$. Therefore in
the absence of interactions between electrons, the only expected
quantum Hall plateaus should occur at every even filling factor
$\nu=2n$. In the experiment of Zhang \emph{et al.}, plateaus at
$\nu=\pm 1$ were observed at large magnetic field, suggesting that
interactions between electrons play a role \cite{Zhang2}. However,
none of the other odd plateaus $\nu=\pm3;\pm5;\dots$ were observed.
It is very simple to understand these facts if one assumes that the
electrons have different on-site energies on the two carbon atoms in
the unit cell, just as for boron nitride, which also has a honeycomb
structure \cite{Semenoff}. In that case the Dirac electrons become
massive \cite{Semenoff} and the LL are almost unchanged except for
the CLL which is not valley degenerate anymore \cite{Haldane}: this
is enough to understand the filling factors at which QH plateaus
were observed. It now remains to find a microscopic mechanism
leading to the generation of this mass. In the following section, we
show that \emph{in the presence of a magnetic field} phonon-mediated
(indirect) interactions between electrons lead to a CDW ground
state. The latter features different electronic charge densities on
the two sublattices providing different on-site energies for A and B
carbon atoms, and therefore leading to a mass for the Dirac
electrons.

\section{Magnetic-field-induced Peierls instability}
\label{MFIPI} Undoped graphene is a very peculiar two-dimensional
metal: it has a Fermi surface consisting of only two points
($\mathbf{K},\mathbf{K'}$), which means that its density of states
at the Fermi energy is vanishingly small. Such a Fermi surface
features perfect nesting at wavevector $\mathbf{Q}=0$ and
$\mathbf{K}$, as testified by the Kohn anomalies found in the
spectrum of its optical in-plane phonons \cite{Piscanec}. This means
that graphene is a quite unstable metal with respect to charge
density modulations at wavevector $\mathbf{Q}$ and therefore,
although two-dimensional, a good candidate for a Peierls instability
\cite{Peierls}. Electrons in free-standing graphene are strongly
coupled to the in-plane optical phonons but are not coupled to the
out-of-plane ZO phonon \cite{Piscanec}. Such a coupling is also
absent in graphite because each graphene layer is in a symmetric
environment with respect to its neighboring layers. However for
graphene on a substrate, this is not the case: as the substrate
(amorphous SiO$_2$) is different from the superstrate (air), the
coupling of electrons to the ZO phonons is not forbidden by
symmetry. Such an electron-phonon coupling is actually provided by
the interactions of the graphene sheet with the substrate, as we
show below. The ZO phonons correspond to an out-of-plane vibration
where the A and B atoms in the unit cell move out of phase. If this
vibration is frozen-in, every second carbon atom is closer to the
substrate than its partner in the unit cell: the honeycomb lattice
is no longer flat but crinkled as an egg box. In that case, the A
and B carbon atoms do not have the same on-site energy anymore,
which means that the Dirac electrons have acquired a finite mass. In
the following, we quantitatively develop this idea.

We now consider a spontaneous out-of-plane lattice distortion which
-- in the presence of the substrate -- breaks the inversion symmetry
of the honeycomb lattice and provides a mechanism for lifting the
valley degeneracy in the CLL. Assume that the A (resp. B) sublattice
moves away (resp. towards) the substrate by a distance $\eta$.
Electrons are still described by a honeycomb nearest-neighbor
tight-binding model, however the two atoms in the basis now have
different on-site energies. The energy on atom A/B is called $\pm
M$. Close to the Dirac points and including the Zeeman effect, the
LL read \cite{Haldane}
\begin{eqnarray}
\varepsilon_{n,\sigma,\alpha}&=&\text{sgn}(n)\sqrt{M^2+2\hbar v_F^2
eB_\perp |n|}
+ \frac{g^*}{2}\mu_B B_\text{tot} \sigma \qquad \text{if} \qquad n \neq 0 \label{LL2}\\
\varepsilon_{0,\sigma,\alpha}&=&\alpha M+ \frac{g^*}{2}\mu_B
B_\text{tot} \sigma \qquad \text{if} \qquad n=0\, , \label{LL3}
\end{eqnarray}
where $\alpha = \pm 1$ is the valley index corresponding to the
Dirac points $\alpha \mathbf{K}$, and $\sigma=\pm 1$ is the spin
projection along the magnetic field direction. In terms of the
low-energy effective theory, the distortion means that the Dirac
electrons spontaneously acquire a finite mass $M/v_F^2$. The on-site
energy difference lifts the valley degeneracy for the CLL
\emph{only}, with a valley gap $\Delta_v=2M$. Note that the effect
of a nonzero on-site energy $M$ on each $n\neq0$ LL is very weak, of
order $M^2/\hbar v_F^2eB_\perp \ll 1$ for a typical magnetic field.
The reason is that in every LL, apart from the CLL, each quantum
state -- independently of its valley index -- has an equal weight on
the two sublattices. This is, however, not the case in the CLL:
electrons at the $K$ (resp. $K'$) point only reside on the $A$
(resp. $B$) sublattice. This explains why breaking the inversion
symmetry of the honeycomb lattice (by introducing different on-site
energies on the two sublattices) lifts the valley degeneracy of the
CLL only.

Such a lattice distortion spontaneously occurs because it lowers the
total energy, just as in Peierls's mechanism \cite{Peierls} except
for the magnetic field playing an essential role here and for the
crystal being two rather than one dimensional. Assume that the last
partially filled LL is $n=0$ (i.e. the gate voltage $V_g$ is such
that $|\nu| \leq 2$). We show that in this case it is always
favorable to slightly distort the lattice \emph{provided} there is a
perpendicular magnetic field \footnote{When $|\nu|>2$, there is no
energy gained by distortion. The distortion only occurs for $\nu$
close to zero.}. The distortion lowers the electronic energy. This
energy lowering comes both from the CLL, which gives an essential
contribution, and also from all the $n<0$ LL, which contribute in a
less important way as we explain below. There are $(2+\nu)N_\phi$
electrons in the CLL. They contribute an energy gain
\begin{equation}
E_{n=0}=-N_\phi (2-|\nu|)M \label{electron}
\end{equation}
because when $\nu<0$, all $(2+\nu)N_\phi$ electrons gain each an
energy $M$ and when $\nu>0$, $2N_\phi$ electrons gain each an energy
$M$ but the remaining $\nu N_\phi$ electrons loose each an energy
$M$. This energy gain depends on the magnetic field through
$N_\phi$. In addition, the energy gain is linear in the out-of-plane
distortion $\eta$ because the on-site energy is proportional to the
distortion, as we discuss below: $M = D \eta$, where $D$ is a
proportionality constant, akin to a deformation potential
\footnote{The corresponding electron-phonon coupling constant
$g_\text{el.-ph.}$ that would appear in a Fr\"ohlich-type
Hamiltonian is $g_\text{el.-ph.}\sim D\sqrt{\hbar/(m_c \omega_0)}$,
where $m_c$ is the mass of a carbon atom.}. The other
$2(N_p-N_\phi)$ electrons that fill the $n<0$ LLs, also contribute
to the energy lowering. Each of them gains a small energy compared
to what an $n=0$ electron gains, as discussed in the preceding
paragraph, but as there are many more of them, about $2(N_p-N_\phi)
\approx 2N_p$ , we cannot neglect their contribution. In the
tight-binding model, we find
\begin{equation}
E_{n<0}=-\gamma \frac{N_p a}{\hbar v_F}M^2 \, , \label{electron2}
\end{equation}
where the numerical factor $\gamma \approx 0.67$. This energy gain
is quadratic in the distortion, and therefore smaller than $E_{n=0}$
at small distortion, and independent of the magnetic field.
Actually, this term represents the full electronic energy gain for a
lattice distortion under \emph{zero} magnetic field. The distortion
costs an elastic energy
\begin{equation}
E_{\text{elastic}}=N_p G \eta^2 \label{elastic}\, ,
\end{equation}
where the out-of-plane distortion is assumed to be small $\eta \ll
a$ and $G$ is an elastic constant, which -- in the adiabatic
approximation -- is related to the ZO phonon frequency at
$\mathbf{q}=0$ by $\omega_0^2\approx 4G/m_c$, where $m_c$ is the
mass of a carbon atom and $\omega_0/2\pi$ is the phonon frequency at
the $\Gamma$ point. From the measured frequency $\omega_0/2\pi c
\sim 800$~cm$^{-1}$ of the \emph{graphite} out-of-plane optical
phonon \cite{Dresselhaus}, we obtain $Ga^2 \approx m_c
\omega_0^2a^2/4 \sim 14$~eV.

The total energy is therefore
\begin{equation}
E_\text{tot}=E_{n=0}+E_{n<0}+E_{\text{elastic}}=-N_\phi
(2-|\nu|)D\eta-\gamma \frac{N_p a}{\hbar v_F}D^2\eta^2+N_p G
\eta^2\, . \label{Etot}
\end{equation}
The most remarkable difference in this expression with respect to
the usual Peierls case is the presence of an energy gain linear in
the lattice distortion $\eta$ rather than proportional to $\eta^2
\ln \eta$ \cite{Gruener}. We note that the phonon mediated
(indirect) electron-electron interaction is
$g_\text{el.-el.}^{\text{eff.}}\sim g_\text{el.-ph.}^2/(\hbar
\omega_0)\sim D^2/G$. The dimensionless electron-electron coupling
constant\footnote{The quantity $\lambda$ is commonly known as the
\emph{dimensionless electron-phonon coupling constant}, e.g. in the
context of the BCS theory of superconductivity, eventhough it is
actually a dimensionless (phonon-mediated) electron-electron
coupling constant \cite{Gruener}.} $\lambda$ is therefore of order
$g_\text{el.-el.}^{\text{eff.}}/t$. Comparing the kinetic energy
gain (in the absence of a magnetic field) and the elastic cost in
equation (\ref{Etot}) shows that it is $\lambda = 2\gamma
D^2/(3Gt)$.

As $E_{n<0}$ and $E_{\text{elastic}}$ are both quadratic in the
lattice distortion, we introduce a renormalized elastic constant
$G'=G-\gamma aD^2/\hbar v_F$ and write an effective elastic energy:
\begin{equation}
E_{\text{elastic}}+E_{n<0}=N_p G' \eta^2 \label{elastic2}\, .
\end{equation}
The effect of the $n<0$ electrons is to reduce the lattice stiffness
and therefore to enhance the distortion. We take it as an
experimental fact that there is no spontaneous out-of-plane
distortion in the absence of a perpendicular magnetic field, see
also \cite{Saito}, which means that $G'>0$. This means that the
Peierls instability does not occur in the absence of a magnetic
field. For the ZO phonon, it means that -- in the adiabatic
approximation -- its frequency is renormalized by the coupling to
the electrons as $\omega^2\approx
\omega_0^2+4(G'-G)/m_c=\omega_0^2(1-\lambda)$. Measuring the ZO
phonon mode in \emph{graphene on substrate} would directly give
access to $G'$, and therefore provide an independent determination
of the constant $D$ through the equation $D=\sqrt{(G-G')\hbar
v_F/\gamma a}$.

In order to obtain the ground state -- within our mean-field
approach -- we minimize the total energy $E_\text{tot}$, see
equation (\ref{Etot}), as a function of the distortion $\eta$ and
get
\begin{equation}
\eta=\frac{N_\phi}{N_p} \frac{2-|\nu|}{2} \frac{D}{G'}\, ,
\label{Peierls}
\end{equation}
from which the valley splitting follows as
\begin{equation}
\Delta_v=\frac{N_\phi}{N_p} (2-|\nu|) \frac{D^2}{G'}\propto B_\perp
\, ,
\end{equation}
and the condensation energy is $E_\text{tot}=-(2-|\nu|)N_\phi
D\eta/2$. Actually in the CLL, and only in that LL, the valley index
is the same as the sublattice index. Therefore if at $\nu\approx 0$
all electrons in the CLL are in one valley, they are also only on
one of the two sublattices: the ground state can therefore be
described as a commensurate A-B (or site centered) CDW. The charge
density oscillation is quite small however: $|N_A-N_B|/N_p \approx
N_\phi/N_p \sim (a/l_B)^2 \ll 1$, where
$l_B=\sqrt{\hbar/(eB_\perp)}$ is the magnetic length.

The usual Peierls instability, which results in a gap opening at the
Fermi energy in the electronic spectrum, is accompanied by a
softening of the coupled phonon mode \cite{Gruener}. Approaching the
instability from the metallic side, the phonon mode softens and goes
to zero precisely at the transition -- a phenomenon known as a giant
Kohn anomaly \cite{Gruener}. In order to see this effect in the case
of the magnetic field induced Peierls instability where the driving
parameter is the magnetic field rather than the temperature, we
consider broadened LL. Broadening is due to disorder and slightly
modifies the preceding calculation for lattice distortion at weak
magnetic field, when the valley splitting $\Delta_v$ is smaller than
the LL width $\Delta_\text{imp}$. For example, for rectangular LL --
the density of states being $4N_\phi/\Delta_\text{imp}$ inside a LL
and zero otherwise -- Eq.~(\ref{electron}) is changed into
\begin{equation}
E_{n=0}=-\frac{2N_\phi}{\Delta_\text{imp}}M^2 =
-\frac{2N_\phi}{\Delta_\text{imp}}D^2 \eta^2 \, ,
\end{equation}
while Eq.~(\ref{electron2}) remains unchanged since it involves
totally filled LLs. The electronic energy gain is now proportional
to $\eta^2$. We therefore introduce still another effective elastic
constant
\begin{equation}
G''=G'-2\frac{N_\phi}{N_p}\frac{D^2}{\Delta_\text{imp}}
\end{equation}
in order to write the total energy as $E_{\text{tot}}=N_p G''
\eta^2$. The renormalized elastic constant $G''$ depends on the
perpendicular magnetic field. A distortion only occurs if $G''< 0$,
i.e. $B_\perp>B_c\equiv 2\pi\hbar
G'\Delta_\text{imp}/3\sqrt{3}ea^2D^2$, which is always the case at
large enough magnetic fields. This condition is precisely equivalent
to requiring that the valley splitting $\Delta_v=2D\eta$ -- given by
Eq.~(\ref{Peierls}) with $\nu \approx 0$ -- be larger than the LL
width $\Delta_\text{imp}$. Therefore, as soon as $B_\perp$ is larger
than the threshold value $B_c$, the lattice is distorted -- in other
words, the ZO phonon is frozen-in -- and the valley gap is larger
than the LL width, which means that one can use the results obtained
in the case of infinitely narrow LL. When $B_\perp<B_c$, the ZO
phonon is strongly renormalized by a magnetic-field-dependent term
and its frequency, in the adiabatic approximation, is
\begin{equation}
\omega(B_\perp)\approx
\omega(0)\left(1-\frac{B_\perp}{B_c}\right)^{1/2}\, ,
\end{equation}
where $\omega(0)\approx \omega_0\sqrt{1-\lambda}$ is the
renormalized phonon frequency at zero magnetic field. The phase
transition toward the CDW ground state occurs at $B_\perp=B_c$,
where $\omega(B_\perp)=0$.

Here we exhibit a microscopic mechanism providing the coupling of
the electrons to the ZO phonon -- i.e. the non-zero deformation
potential $D$ -- via the interaction of the carbon atoms to the
substrate. It is quite difficult to accurately predict the constant
$D$ and we will therefore only provide an order of magnitude
estimate. The mechanism that we think gives the largest contribution
results from the interaction of a single carbon atom from the
graphene sheet with the amorphous SiO$_2$ substrate treated as a
dielectric continuum. The non-retarded Lennard-Jones interaction
energy of an atom at a distance $r$ of a dielectric wall is given by
$E_\text{LJ}(r)\approx -(\epsilon_r-1)\langle
\mathbf{d}^2\rangle/(\epsilon_r+1)48\pi \epsilon_0 r^3$, where
$\langle \mathbf{d}^2\rangle$ is the atomic ground state expectation
value of the squared electric dipole moment \cite{Aspect}. The
on-site energy change resulting from the lattice distortion may be
estimated as
\begin{equation}
\pm M \approx E_\text{LJ}(d_0\pm \eta)-E_\text{LJ}(d_0)\approx \pm
\frac{\epsilon_r-1}{\epsilon_r+1}\frac{\langle
\mathbf{d}^2\rangle}{16\pi \epsilon_0 d_0^4} \eta
\end{equation}
where the $\pm$ sign refers to sublattice A ($+1$) or B ($-1$),
$d_0$ is the average distance between the graphene sheet and the
substrate and we assumed that $\eta \ll d_0$. For a carbon atom
$\sqrt{\langle \mathbf{d}^2\rangle} \sim 4ea_0$, where $a_0$ is the
Bohr radius, which gives $Da\sim
a(\epsilon_r-1)e^2a_0^2/(\epsilon_r+1)\pi\epsilon_0 d_0^4\sim 1$ to
$14$~eV depending on $d_0\sim$ $0.1$ to $0.2$~nm. Therefore, the
order of magnitude of the deformation potential $Da$ is $5$~eV. The
condition $G'>0$ that no Peierls instability occurs when $B_\perp=0$
implies that $Da<\sqrt{Ga\hbar v_F /\gamma}\approx 9.8$~eV.
Therefore, in order to match the experiment \cite{Zhang2}, we take
the plausible value $Da=7.8$~eV, which gives $G'a^2\approx 4.2$~eV.
In the end, the dimensionless electron-phonon coupling constant
$\lambda= 2\gamma D^2/(3Gt) \approx 0.8$, which is quite a large
value.

The mechanism proposed here features several differences with
respect to the standard Peierls instability \cite{Peierls,Gruener}:
(i) the metal is two- rather than one-dimensional; (ii) the
distortion does not change the Bravais lattice but only the point
group symmetry -- the inversion symmetry of the honeycomb crystal is
broken by the instability but the triangular Bravais lattice is not
affected by the distortion; (iii) the electron-phonon coupling does
not exist in the isolated graphene crystal but is provided via the
coupling to the substrate; (iv) the control parameter for the
instability is the magnetic field rather than the temperature; (v)
the metal (undoped graphene) is actually a zero gap semiconductor
and has $k_F=0$ and the nesting condition is $|\mathbf{Q}|=0$, which
indeed equals $2k_F$.

\section{Charge density wave in graphene: direct electron-electron interactions}
\label{CDWgraphene} Many authors have considered the effect of
direct (Coulomb) electron-electron interactions in graphene, in
particular to study the quantum Hall effect. Depending on which
aspect of the Coulomb interaction is thought to dominate, different
scenarios have been proposed. A popular scenario is that of quantum
Hall valley and spin ferromagnetism, see e.g.
\cite{Alicea,Nomura,Abanin,Fertig,Goerbig}. This is the graphene
version of the well-known quantum Hall ferromagnetism: the exchange
interaction between electrons favors ferromagnetism inside a
partially filled LL. Indeed, it allows one to reduce the interaction
energy -- in other words, to gain exchange energy -- at no kinetic
energy cost when the LL are assumed infinitely narrow and flat. This
mechanism predicts plateaus at every integer filling factor in
contradiction with the existing experiments. A possible way out of
this contradiction is to consider strong disorder, which naively
speaking, broadens the LL and therefore, by restoring the
possibility of kinetic energy changes, works against ferromagnetism
\cite{Nomura}. This could explain the absence of certain QH plateaus
at magnetic fields up to $45$~T.

Apart from QH ferromagnetism, we are aware of only one other
scenario: namely that of a commensurate A-B CDW ground state
triggered by the magnetic field. This scenario exists in three
different microscopic versions depending on which electron-electron
interaction is believed to dominate. One version -- the magnetic
field induced Peierls instability -- is the subject of the present
paper and relies on electron-phonon coupling providing indirect
interactions between electrons. The other two rely on Coulomb
(direct) electron-electron interactions: one -- the so-called
magnetic catalysis -- focusses on long-range Coulomb interactions,
while the other relies on lattice-scale (extended Hubbard type)
Coulomb interactions. All of these mechanisms lead to a similar
ground state and therefore predict the same QH plateaus.

\subsection{Long range Coulomb interactions}
In the context of highly oriented pyrolytic graphite, Khveshchenko
suggested that long range Coulomb interactions could also lead to a
CDW ground state in the presence of a magnetic field
\cite{Khveshchenko}. This idea stemmed from the phenomenon of
dynamical chiral symmetry breaking originally proposed in
relativistic theories of (2+1)-dimensional interacting Dirac
fermions. In this scenario, the electron-electron interaction comes
from the Coulomb interaction in the continuum limit in which only
the long range part of the interaction survives. Within the CLL,
electrons and holes pair up to form excitons, as a result of the
attractive Coulomb interaction, and these excitonic pairs Bose
condense. This leads to the creation of an excitonic gap in the
electronic spectrum, which is similar to what we called the valley
gap $\Delta_v$, but has a different magnetic-field dependence
$\Delta_\text{exciton}\sim e^2/(\epsilon l_B) \propto
\sqrt{B_\perp}$ \footnote{This excitonic gap is of the same order of
magnitude as the LL spacing $\hbar \omega_c$ as the ``graphene fine
structure constant" $e^2/(4\pi \epsilon \hbar v_F)\approx
 0.5$.}. This mechanism only occurs in the presence of a magnetic field
and has therefore been dubbed the ``magnetic catalysis'' of the
excitonic instability \cite{Khveshchenko,Gusynin}. The ground state
can be described either as a Bose condensate of excitons inside the
CLL or as an A-B CDW.

\subsection{Lattice-scale Coulomb interactions}
Several years ago, in the context of STM studies on graphite
surfaces, Tchougreeff and Hoffmann \cite{Hoffmann,Wagner} suggested
that graphene could have a CDW ground state \footnote{Their work was
actually triggered by an early experiment on a \emph{graphene} sheet
on the Pt(111) surface \cite{Land}}. They considered Coulomb
interactions between electrons in an extended Hubbard model on the
honeycomb lattice with on-site repulsion $U \sim e^2/(\epsilon a)$
and nearest-neighbor repulsion $V \sim U$, and neglected the
long-range part of the Coulomb interaction. They showed that if the
nearest-neighbor repulsion $V$ was strong enough to overcome the
effect of the on-site repulsion $U$, a CDW ground state should be
favored. In the opposite case, a spin density wave (SDW) would
result. Recently, several authors have shown that a magnetic field
should help to trigger this CDW instability \cite{Alicea,Herbut} --
or the SDW depending on the precise $U/V$ ratio \cite{Herbut}. In
the CDW case, the corresponding valley gap is linear in the magnetic
field $\Delta_\text{Hubbard}\sim U (a/l_B)^2 \propto B_\perp$ just
as in the magnetic-field-dependent Peierls mechanism $\Delta_v \sim
(D^2/G')(a/l_B)^2$.

\section{Conclusion}
\label{Conclusion} We have reviewed the proposal of a
magnetic-field-induced Peierls instability in graphene leading to a
CDW ground state \cite{Fuchs}. The driving mechanism is the electron
- ZO phonon coupling, which is provided via the interaction of
carbon atoms to the substrate. The instability corresponds to the
spontaneous breaking of the inversion symmetry of the graphene
honeycomb lattice and results in Dirac electrons having an effective
mass proportional to the magnetic field. Apart from electron-phonon
interactions, there are other interactions leading to a CDW
instability triggered by a perpendicular magnetic field
\cite{Khveshchenko,Gusynin,Alicea,Herbut}. As to the question of
which is the dominant interaction leading to the instability in
graphene, we mention two related examples of one-dimensional metals
featuring a CDW instability. On the one hand, consider the
well-studied case of polyacetylene, which dimerizes at low
temperature and has long been described as being the prototypical
example of a Peierls instability driven by electron-phonon
interactions. It is now known that Coulomb interactions also play a
very important role in the instability and that electron-phonon
coupling alone is not enough to quantitatively explain the measured
effect \cite{Koenig}. On the other hand, in a one-dimensional metal
like KCP, the dominant mechanism leading to the Peierls distortion
has been shown to be electron-phonon interactions \cite{Comes}. In
the case of graphene on substrate, it is therefore not clear at the
moment whether a CDW does occur and if so, which is the driving
mechanism leading to such a ground state in the presence of a
magnetic field.

\section*{Acknowledgments}
We thank M.O. Goerbig for useful discussions, and I. Herbut and D.
Khveshchenko for comments on the manuscript.

%\begin{figure}
%\resizebox{0.75\columnwidth}{!}{%
%  \includegraphics{fig1.eps} }
%\caption{Please write your figure caption here.}
%\end{figure}
%\begin{table}
%\caption{Please write your table caption here.}
%\begin{tabular}{lll}
%\hline\noalign{\smallskip}
%\end{tabular}
%\end{table}
%

\end{document}